\documentclass[11pt]{article}
\usepackage{graphicx}
\usepackage{cancel}[]
\usepackage{amssymb,amsmath}
\usepackage{ifpdf}
\usepackage{subfigure}
\usepackage{hyperref}[]
\usepackage{fullpage}
\usepackage{color}
\usepackage{ulem}
\ifpdf
    \DeclareGraphicsExtensions{.pdf}
\else
    \DeclareGraphicsExtensions{.eps}
\fi

\begin{document}

\vspace{0.1 in}
\centerline{\Large{\bf{The Sagnac effect and the Tevatron }}}

\vspace{0.1 in}
\centerline{\rm{ A. C. Melissinos}}

\centerline{1/15/2011}

\vspace{0.25 in}

The rotation of the Earth modifies the path-length of particles moving
in opposite directions on the same closed trajectory. We designate the
area enclosed by the trajectory by $\bf{\rm{A}}$, where the vector
indicates the normal to the trajectory plane; the Earths's
angular momentum is ${\bf\Omega}$. The time difference between the two
particles is
\begin{equation}
\delta t = \frac{4\ \bf{A}\cdot \bf{\Omega}}{v^2}
\end{equation}
with $v$ the velocity of the particles. This effect, predicted by
Sagnac \cite{Sagnac} in 1913 , was observed by Michelson \cite{Michelson}
in 1925, using an optical interferometer. In this case when the
two beams merge after a single turn, their relative phase is shifted by
\begin{equation}
\delta \phi = \frac{2\pi}{\lambda}\frac{4\ \bf{A}\cdot \bf{\Omega}}{c}
\end{equation}
where $\lambda$ is the wavelength of the light.

Today's laser gyros consist of a ring cavity in which a lasing medium
is inserted. Due to the Earth's (or the ring's) rotation the two
counter-propagating beams see different cavity lengths and thus have different
frequencies. One measures the beat frequency $\delta f$
between the two oppositely circulating beams,
\begin{equation}
\delta f = \frac{4\ \bf{A}\cdot \bf{\Omega}}{\lambda \ P}
\end{equation}
with $P$ the perimeter of the closed trajectory. In modern ring
interferometers \cite{Stedman} $\delta f$ can be measured to $\sim 1\ \mu$Hz,
and thus the Earth rotation frequency to 1 part in $10^7$.\\

The Sagnac effect should be present in the Tevatron, with the important
difference that the trajectory is determined by the machine lattice and
the particle energy, and by suitable feedback, the two beams are forced to
collide at the intersection regions. Let us consider for the moment single
bunches of protons and antiprotons and assume that both follow the same orbit
and that their momenta are exactly equal, since this determines the mean
orbit radius. If the two bunches are coincident at some point on their
trajectory, then after one turn they will be separated at that point by
a distance $\delta s$,
\begin{equation}
\delta s =  4|\bf{A}||\bf{\Omega}| \rm{sin}\phi /c = 2\times 10^{-6} \rm{m},
\end{equation}
\noindent where $\phi = 41.83^{\circ}$ is the Fermilab latitude, $|{\bf{A}}| = \pi \times
10^6\ \rm{m}^2$ and $|{\bf{\Omega}}|= 2\pi / 86,400 = 7.27 \times 10^{-5}$ r/s
is the diurnal earth rotation angular frequency. Since the beam rotation
frequency is $f \sim 50$ kHz, in one minute the two bunches should be
separated by $\sim 6$ m.\\

The problem with this proposal is that it assumes that both bunches are
coasting at exactly the same mean radius. The required precision is
\begin{equation}
\frac{\delta R}{R} < \frac{\delta s}{2 \pi R} \sim \frac{10^{-9}}{\pi}.
\end{equation}
Even accounting for the momentum compaction in strong focussing machines,
\begin{equation}
\frac{dp}{p} = Q^2 \frac{dR}{R},
\end{equation}
where $Q$ is the betatron tune ($Q \sim 50$), the relative momentum difference
between the proton and antiproton bunch must be $\sim 10^{-6}$ which is
unattainable. Furthermore, any shift in the bunch longitudinal position
affects the rf phase which will restore the centroid of the bunch to the
center of the rf bucket. The terrestrial distortion of the machine lattice
would affect both bunches equally.\\

{\bf{Demonstration of the Sagnac formula}}\

 Consider a ring of radius $r$
on a plane surface that is rotating with angular velocity $\Omega$. A particle
of velocity $v$ is moving on the ring in the direction of rotation. The
time required for a complete turn is
\begin{equation}
t_{+} = \frac{2\pi r}{v} + \frac{\Omega r t_{+}}{v}\qquad {\rm{or}}\qquad
t_{+} = \frac{2\pi r}{v - \Omega r} \nonumber
\end{equation}
For a particle moving opposite to the rotation direction
\begin{equation}
t_{-} =  \frac{2\pi r}{v + \Omega r}\qquad {\rm{and}} \qquad
\delta t = t_{+} - t_{-}
= \frac{4\pi r^2 \Omega}{v^2 -(\Omega r)^2}
\end{equation}
The projection of ${\bf{\Omega}}$ on the normal to the ring's plane is
${\bf{\Omega \cdot A}}$, $A = \pi r^2$ and we can always assume $\Omega r \ll v$.\\

{\bf{Relativistic derivation}}\

We obtain the same result by considering the increase/decrease in the
velocity of the $p/\overline{p}$ bunches. If the linear velocity of the
ring is $\beta_r$, the velocity difference $d\beta = 2 \beta_r/2 \gamma^2$,
where $\beta ,\gamma$ refer to the $p/\overline{p}$. It follows that
\begin{equation}
\frac{dp}{p} = \frac{d\gamma}{\gamma}= 2 \beta_r =
\frac{2\Omega\ {\rm{sin}}\phi\ R}{c} = \frac{\delta s}{2\pi R},
\end{equation}
as in Eq. 9.

\end{document}